\documentclass{optica-article}

\journal{opticajournal} 

\articletype{Research Article}

\usepackage{lineno}

\usepackage{graphicx}  
\usepackage{subfigure}
\usepackage{subfigure}
\usepackage{multirow}
\usepackage{appendix}
\usepackage{float}
\usepackage{xcolor}
\usepackage{tcolorbox}

\begin{document}

\title{Triple-Tone Microwave Control for Sensitivity Optimization in Compact Ensemble Nitrogen-Vacancy Magnetometers}

\author{Ankita Chakravarty,\authormark{1,2,*} Romain Ruhlmann,\authormark{5} Vincent Halde,\authormark{5} David Roy-Guay,\authormark{5} Michel Pioro-Ladrière,\authormark{1,2} Lilian Childress,\authormark{4} and Yves Bérubé-Lauzière\authormark{1,3}}

\address{\authormark{1}Institut quantique, Université de Sherbrooke, Sherbrooke, Québec J1K 2R1, Canada\\
\authormark{2}Département de physique, Université de Sherbrooke, Sherbrooke, Québec J1K 2R1, Canada\\
\authormark{3} Département de génie électrique et de génie informatique, Université de Sherbrooke, Sherbrooke, Québec J1K 2R1, Canada\\
\authormark{4}Department of Physics, McGill University, 3600 Rue University, Montréal, Québec H3A 2T8, Canada\\
\authormark{5}SBQuantum, 805 Rue Galt O, Sherbrooke, Québec J1H 1Z1, Canada}

\email{\authormark{*}Yves.Berube-Lauziere@USherbrooke.ca} 


\begin{abstract*}
Ensembles of nitrogen-vacancy (NV) centers in diamond are a well-established platform for quantum magnetometry under ambient conditions. One challenge arises from the hyperfine structure of the NV, which, for the common $^{14}$N isotope, results in a threefold reduction of contrast and thus sensitivity. By addressing each of the NV hyperfine transitions individually, triple-tone microwave (MW) control can mitigate this sensitivity loss.   Here, we experimentally and theoretically investigate the regimes in which triple-tone excitation offers an advantage over standard single-tone MW control for two DC magnetometry protocols: pulsed optically detected magnetic resonance (ODMR) and Ramsey interferometry. 
We validate a master equation model of the NV dynamics against ensemble NV measurements, and use the model to explore triple-tone vs single-tone sensitivity for different MW powers and NV dephasing rates. For pulsed ODMR, triple-tone driving improves sensitivity by up to a factor of three in the low-dephasing regime, 
with diminishing gains when dephasing rates approach the hyperfine splitting. In contrast, for Ramsey interferometry, triple-tone excitation only improves sensitivity if MW power is limited. 
Our results delineate the 
operating regimes where triple-tone control provides a practical strategy for enhancing NV ensemble magnetometry in portable and power-limited sensors 
\end{abstract*}


\section{Introduction}
\label{sec:introduction}
Nitrogen-vacancy (NV) centers in diamond have emerged as a versatile platform for quantum sensing \cite{Degen_2017} and information processing \cite{Xu_2019}, owing to their spin-dependent fluorescence, long coherence times, and microwave addressability under ambient conditions \cite{Doherty_2013, Dobrovitski_2013}. A key application is 
magnetometry \cite{Rondin_2014,Taylor_2008}, 
where NV sensors have driven advances in biology \cite{Schirhagl2014,Aslam2023}, geophysics \cite{Glenn_2017}, and material science \cite{Thiel_2016}. While single NVs can offer nanoscale resolution, NV ensembles enable higher signal-to-noise ratios 
with micron-scale wide-field imaging capabilities \cite{Boretti2019}, making NV magnetometry attractive for both research and field applications. 

DC magnetometry with NV centers \cite{Degen_2017,Barry_2020} detects static magnetic fields with high sensitivity by probing the NV spin transition frequencies with resonant microwaves (MW). In the laboratory, performance is limited by spin density and coherence, along with fluorescence contrast and collection efficiency \cite{Barry_2020}. 
Field-deployable systems face additional constraints on MW delivery and sequence complexity, motivating simple protocols that maximize sensitivity under realistic conditions. One ubiquitous issue is the hyperfine structure of the NV, which for the common $^{14}$N isotope triples the number of resonance lines while reducing their contrast by a factor of three. 
Nuclear spin polarization \cite{B_rgler_2023,PhysRevB.81.035205, PhysRevResearch.2.023094,Huillery_2021,PhysRevB.80.241204,Fischer_2013} 
can mitigate this contrast loss, 
but requires additional experimental overhead, limiting its practicality. A simpler alternative is to drive all three nitrogen hyperfine transitions simultaneously using a triple-tone MW field.  By simultaneously driving all three hyperfine transitions such that each tone matches a hyperfine transition, the resulting fluorescence contributions add constructively, increasing signal contrast and enhancing magnetic sensitivity. This method requires no extra optical or RF hardware and can be implemented via frequency modulation, making it appealing for compact systems.  

Triple-tone  excitation has been employed in many experiments to date \cite{Barry2020,Ahmadi_2017, Poulsen_2022,zhang2022,Mariani2022,Graham_2023, Webb2019, Arai_2022, Li2018}, however, it does not always provide the ideal threefold sensitivity gain, opening the question of its regimes of applicability.  While triple-tone excitation has been employed in continuous-wave (CW) ODMR as well as pulsed protocols,  we focus here on pulsed approaches for two reasons.  First, the impact of inhomogeneous optical excitation on CW ODMR reduces the universality of results, as magnetometer performance depends on the details of optical illumination as well as MW excitation; secondly, pulsed ODMR and Ramsey offer improved sensitivity relative to CW ODMR by eliminating optical broadening.  In this work, we therefore investigate how triple-tone MW driving affects the sensitivity of pulsed ODMR and Ramsey measurements in NV ensembles. We develop simulations that model the system dynamics, which are validated with our experimental data and 
used to compare single- and triple-tone control for different MW powers 
and spin dephasing rates. Sensitivity is quantified using two metrics: the signal slope, that is, the change in signal with respect to the MW probe frequency (which has the same effect as magnetic field), which is inversely proportional to sensitivity when initialization and readout times are long, and the slope normalized by the square root of the sequence duration, which scales inversely with sensitivity when overhead time is negligible. The higher these metrics, the better the sensitivity. Our results show that triple-tone driving yields substantial sensitivity gains for pulsed ODMR in the low dephasing regime. In Ramsey interferometry, however, such improvements appear only 
when MW power is limited. Comparing both protocols, we find nuanced trade-offs that depend on the specific operating regime. These findings clarify when multi-tone control is advantageous and provide guidance for practical NV-based DC magnetometry.

The paper is organized as follows. Section~\ref{sec:theoretical_background} reviews NV magnetometry and relevant pulse sequences. Section~\ref{sec:experimental_details} describes the experimental setup and workflow. Sections~\ref{sec:pulsed_odmr} and \ref{sec:ramsey} present results for pulsed ODMR and Ramsey interferometry respectively. Section~\ref{sec:discussion} compares their performance, and Section~\ref{sec:conclusion} concludes with key findings and outlook.

\section{Theoretical Background}
\label{sec:theoretical_background}
The negatively charged nitrogen-vacancy (NV) center in diamond is a spin-1 defect, consisting of a substitutional nitrogen atom adjacent to a lattice vacancy \cite{Doherty_2013}. It can be optically initialized and read out via spin-dependent fluorescence, making it a widely used system in quantum sensing. Under green laser excitation (typically at 532 nm), the NV spin undergoes a spin-selective intersystem crossing, allowing the spin projection along the NV axis to be inferred from differences in emitted red fluorescence. Together with MW control of the spin-triplet ground state, this allows coherent spin manipulation for magnetometry and other sensing applications \cite{Degen_2017}.

The ground-state Hamiltonian of the $NV^-$ center \cite{Barry_2020}, is given by
\begin{equation}
\begin{split}
\frac{H}{h} = 
\Delta S_z^2 + \gamma_e (\mathbf{B} \cdot \mathbf{S})
\;+\;
\mathbf{S} \cdot \mathbf{A} \cdot \mathbf{I}
+ P I_z^2
- \gamma_n (\mathbf{B} \cdot \mathbf{I}),
\end{split}
\label{eq:hamiltonian_NV}
\end{equation}
where $\Delta \approx 2.87$~ GHz 
is the zero-field splitting, 
$\gamma_e B_z S_z$ represents the Zeeman interaction with an external magnetic field $B_z$, $\mathbf{S} \cdot \mathbf{A} \cdot \mathbf{I}$ describes the hyperfine coupling between the NV electron spin and the host $^{14}$N nuclear spin, $P$ is the nuclear electric quadrupole moment and the term associated with it is the nuclear quadrupole term, and $\gamma_n$ is the nuclear gyromagnetic ratio. The nuclear hyperfine tensor is given by $\mathbf{A} = \mathrm{diag}(A_\perp,\,A_\perp,\,A_{\parallel})$; here we neglect transverse terms $A_{\perp}$ (see Appendix \ref{appendix:theory}) and the axial component $A_{\parallel}$ splits each spin transition into three components separated by approximately 2.16~MHz \cite{PhysRevB.81.035205}.

We focus on DC magnetometry with NV ensembles \cite{Barry_2020, Degen_2017}, using pulsed optically detected magnetic resonance (ODMR) and Ramsey interferometry (Fig. \ref{fig:experimental_details}(c)). In pulsed ODMR, a resonant $\pi$-pulse is applied after the optical initialization, and the resulting spin population transfer is optically read out via fluorescence. In Ramsey interferometry, a $\pi/2$ pulse precedes the free evolution time $\tau$, followed by a second $\pi/2$ pulse that is phase-shifted by $\pi/2$ relative to the first. During the free evolution interval, the spin accumulates phase, which the second pulse maps into a population difference that is optically read out. In both cases, the measured 
fluorescence signal $C$ 
depends on the spin population difference induced by the MW pulses. The sensitivity \cite{Barry_2020} of the magnetometer is commonly quantified as:
\begin{equation}
\eta = \Delta B \sqrt{T_\text{tot}},
\label{eq:sensitivity}
\end{equation}
where $\Delta B$ is the smallest resolvable magnetic field and $T_\text{tot}$ is the total measurement time, including initialization, control, and readout. Equivalently, using the fluorescence signal, sensitivity can be expressed as
\begin{equation}
\eta = \frac{\Delta C}{(\frac{dC}{d\nu} \frac{d\nu}{dB})} \sqrt{T_\text{tot}},
\label{eq:sensitivity2}
\end{equation}
where $\Delta C$ is the standard deviation in $C$ representing the photon-shot-noise–limited uncertainty, $dC/d\nu$ is the slope of the measured signal with respect to the MW frequency $\nu$ and $d\nu/dB$ depends on the orientation of the magnetic field with respect to the NV axis. Since we pick a single NV orientation to do all the experiments, $d\nu/dB$ remains constant for all the experiments. We focus on comparing two metrics for the sensitivity: the slope $dC/d\nu$ and slope$/\sqrt{T}$ 
where $T$ is the MW manipulation time and $T_\text{tot} = T + T_\text{overhead}$. If the readout noise (and thus $\Delta C$) is identical across protocols, these two metrics are inversely proportional to the sensitivity in the limit of large (slope) and small (slope/$\sqrt{T}$) overhead times, whereby for large overhead times, $T_\text{tot} \approx  T_\text{overhead}$ and for small overhead times $T_\text{tot} \approx  T$. 
The sensitivity comparison thus reflects the intrinsic advantage of single- versus triple-tone driving.

To complement experiments, we perform numerical simulations by solving the Lindblad master equation for an effective spin-$\tfrac{1}{2}$ system with pure dephasing \cite{poulsen2021} (Appendix~\ref{appendix:theory}). While this model neglects ensemble inhomogeneity, 
it reproduces the key features of pulsed ODMR and Ramsey data at low and moderate MW powers. This model enables us to map sensitivity landscapes under single and triple-tone driving, assess performance across dephasing regimes and MW power, and identify optimal conditions for NV ensemble magnetometry.


\section{Experimental Details}
\label{sec:experimental_details}
\begin{figure}[htb]
  \centering
    \includegraphics[width=0.8\textwidth]
    {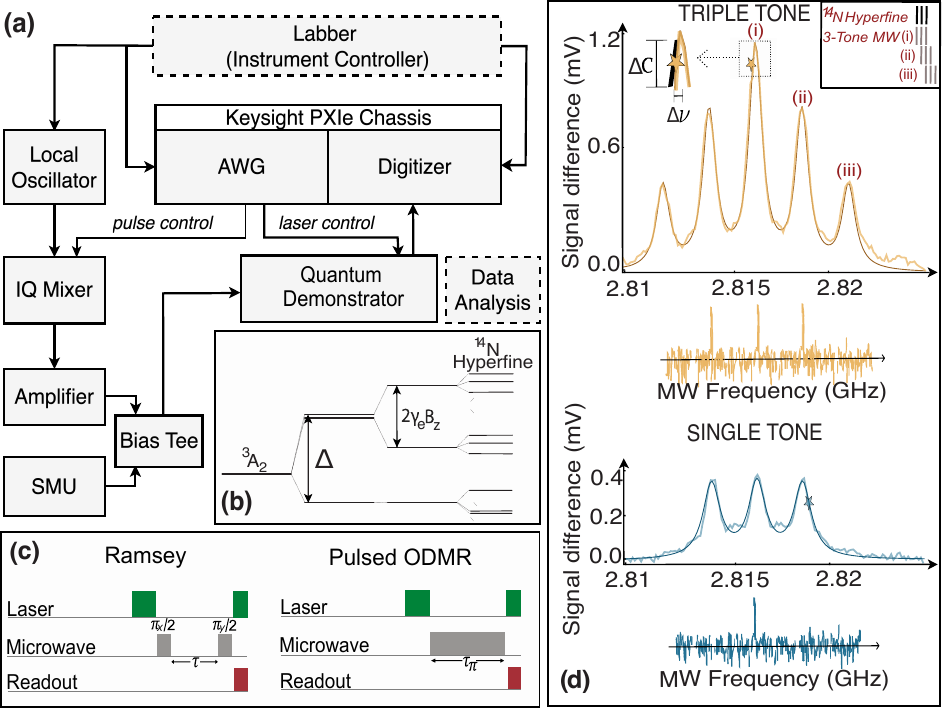}
  \caption{ (a) Schematic of the experimental setup for implementing MW control: An AWG produces the signal, which is mixed with the local oscillator in the IQ mixer. The output is amplified and delivered to the NV ensemble inside the Quantum Demonstrator, which houses the magnetometer that integrates optical excitation and microwave delivery. Fluorescence is collected and digitized using a high-speed digitizer, enabling measurement of spin-state-dependent signals. (b) Ground-state energy level diagram of the negatively charged nitrogen-vacancy (NV) center in diamond, showing the spin-1 triplet sublevels and hyperfine splitting from the host $^{14}$N nuclear spin. The $m_s = 0 \leftrightarrow \pm1$ transitions each split into three lines spaced by 2.16~MHz. (c) Schematic of pulsed ODMR (right) and Ramsey (left) sequences illustrating laser and MW pulse timings. (d) Illustration of single-tone (bottom) and triple-tone (top) MW excitation, where the latter addresses all hyperfine transitions simultaneously, enhancing fluorescence contrast; five resonance features arise due to the overlap of the three tones with the three $^{14}$N hyperfine transitions as shown in the top right inset (case (i): all three tones resonant with hyperfine transitions; case (ii): two resonant tones; case (i): one resonant tone). The ODMR curves are real data where the measured signal is plotted as a function of the MW frequency, and the tones sent are a cartoon representation.}
  \label{fig:experimental_details}
\end{figure}

MW control and signal acquisition were implemented using a fully integrated NV magnetometry system referred to as the Quantum Demonstrator in Fig. \ref{fig:experimental_details}(a), where the diamond sample and optical components are housed in this compact module developed by SBQuantum \cite{halde2025}. This module gives the flexibility to test out MW pulse control schemes on the magnetometer. Control and readout were carried out using modular Keysight PXIe-based hardware, providing synchronized MW pulse generation and fluorescence detection with nanosecond precision.

Figure~\ref{fig:experimental_details}(a) shows a schematic of the setup, illustrating the full system from MW pulse generation to signal detection. MW pulses were synthesized using an arbitrary waveform generator (AWG; Keysight M3202A) in a PXIe chassis (Keysight M9019A). The AWG generated in-phase (I) and quadrature (Q) signals such that
$V_\text{AWG}(t) = I(t)\cos(2\pi f_\text{IF}t) + Q(t)\sin(2\pi f_\text{IF}t)$,
with $f_\text{IF} = 100~\text{MHz}$. These signals were mixed with a local oscillator (LO; Keysight PSG E8275D) in an IQ mixer (Keysight U3022A H37) to generate the final MW drive at $f_\text{MW} = f_\text{LO} + f_\text{IF}$. 
For triple-tone excitation, we directly synthesize a superposition of three tones with the AWG, controlling the pulse amplitude, duration, phase, and frequency for each tone.

After mixing, the MW signal was amplified and delivered to the Quantum Demonstrator, which houses the magnetometer that integrates optical excitation and microwave delivery. The system includes an internal 520~nm laser diode (maximum output power 110~mW) and a dual-post re-entrant microwave cavity. Both the laser and MW drive can be externally controlled, allowing for flexible experimental configurations while maintaining robust performance. The diamond sample (Element Six, DNV-B1), mounted inside the cavity, is a $1\times1\times0.5~\mathrm{mm}^3$ single-crystal diamond with an NV concentration of 300~ppb and a $T_2^*$ coherence time of 1~$\mu$s. A static magnetic field of 2~mT was applied along the NV axis, lifting the degeneracy of the $m_s = \pm 1$ states. A plano-convex lens with a focal length of 15~mm and a diameter of 6~mm was used to focus the optical excitation, such that the beam waist was positioned near the center of the 0.5~mm-thick diamond.

Spin-state-dependent fluorescence was collected with an estimated 1\% collection efficiency and digitized using a high-speed digitizer (Keysight M3102A). Each sequence alternated between MW-on and MW-off conditions, and the spin signal was obtained from their difference \cite{Misonou_2020} (referred to as the signal difference in all experimental data). IQ modulation enabled direct generation of multi-tone pulses, including the triple-tone excitation addressing all three hyperfine transitions simultaneously, without additional hardware. This combination of integrated optics, microwave control, and engineered diamond enables reproducible ensemble-based NV measurements within a compact footprint.

For our apparatus, the signal slope ($dC/d\nu$) provides the most relevant sensitivity metric, as the preparation and readout times are much longer than the control sequence (200~µs window vs µs-scale MW pulses). Nevertheless, we also examine slope$/\sqrt{T}$, which would apply for faster initialization/readout and/or longer pulse sequences,  extending the applicability of our investigation to a broader range of possible experiments. 


\section{Pulsed ODMR}
\label{sec:pulsed_odmr}
We next evaluate the performance of triple-tone MW driving using pulsed optically detected magnetic resonance (ODMR) \cite{Dreau_2011}, implemented with the setup described in Section~\ref{sec:experimental_details}. Pulsed ODMR enhances sensitivity by replacing continuous-wave driving with short MW pulses, suppressing optical power broadening and yielding narrower linewidths and steeper slopes \cite{Barry_2020, Wang2024}. 
Triple-tone excitation enhances ODMR contrast—up to threefold in the ideal case. The ODMR spectra for single- and triple-tone protocols are fit to Lorentzian model profiles: three peaks for the single-tone case and five for the triple-tone case. The NV ground state contains three hyperfine transitions associated with the $^{14}$N nucleus. Triple-tone MW driving produces five observable resonance features because of how each tone interacts with all three transitions depending on the detuning (Fig.~\ref{fig:experimental_details}(d)) : (i) a central peak where all three tones coincide with the three hyperfine transitions, giving roughly triple contrast; (ii) two intermediate peaks where two tones overlap with two transitions, producing approximately double contrast; and (iii) two outer peaks where only a single tone is resonant, yielding single-tone contrast. From these three- and five-Lorentzian fits, we extract the maximum slope (Fig.~\ref{fig:experimental_details}(d)) to determine the sensitivity metrics for single- and triple- tone excitation.

\begin{figure}[htb]
  \centering
    \includegraphics[width=0.8\textwidth]
    {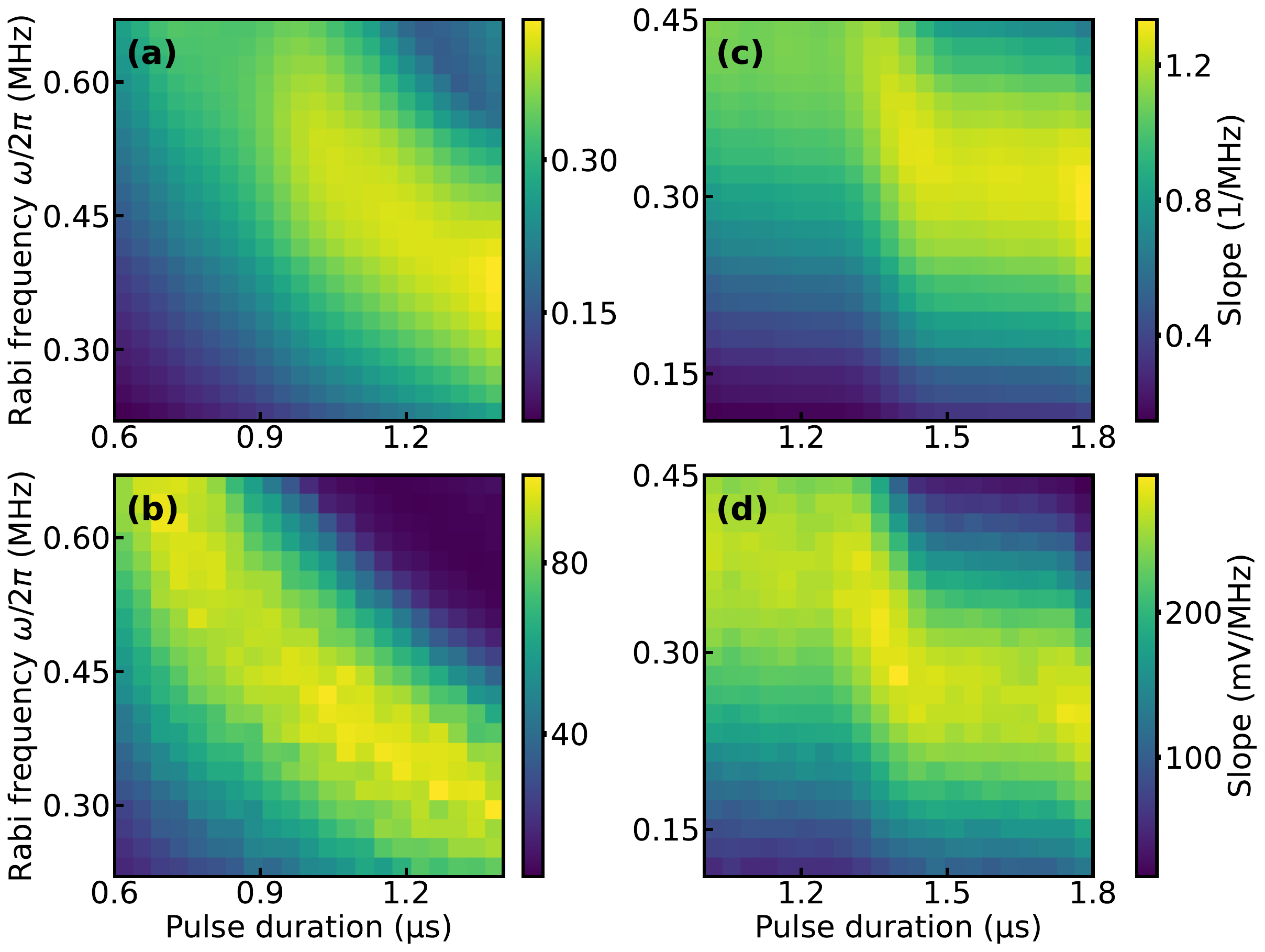}
  \caption{Pulsed ODMR slope maps for single-tone (a-b) and triple-tone (c-d) driving. (a, c) Simulated slopes as functions of Rabi frequency and pulse duration. (b, d) Corresponding experimental data acquired on a 21×21 grid of pulse durations and MW amplitudes. 
  }
  \label{fig:pulsed_odmr_combined_final_2}
\end{figure}

For both single- and triple-tone excitation, we performed two-dimensional sweeps of the MW pulse duration and
applied MW power.
Figure~\ref{fig:pulsed_odmr_combined_final_2} compares experimental ODMR maps with simulations. In (b) and (d), the experimental Rabi frequencies are approximated from the MW voltage amplitude using a 2.24 MHz/mV calibration based on Rabi oscillations taken 
at the same time as the data. However, we later observed fluctuations in the voltage-to-Rabi-frequency zero offset of $\sim$100 kHz, which likely underlies the small vertical offsets between the simulation and experiment. A vertical step visible in the triple-tone data arises from driving detuned hyperfine transitions. Extracted sensitivities confirm enhancement: for the slope, we measure 
a single-tone to triple-tone sensitivity ratio of $2.93 \pm 0.09$, in excellent agreement with the simulated ratio of 2.93. Slope$/\sqrt{T}$ 
yields a sensitivity ratio of $2.28 \pm 0.05$, slightly below the simulated 2.49, likely due to the maximum occurring at the edge of the measured and simulated parameter range.

\begin{figure}[htb]
  \centering
    \includegraphics[width=1.0
    \textwidth]{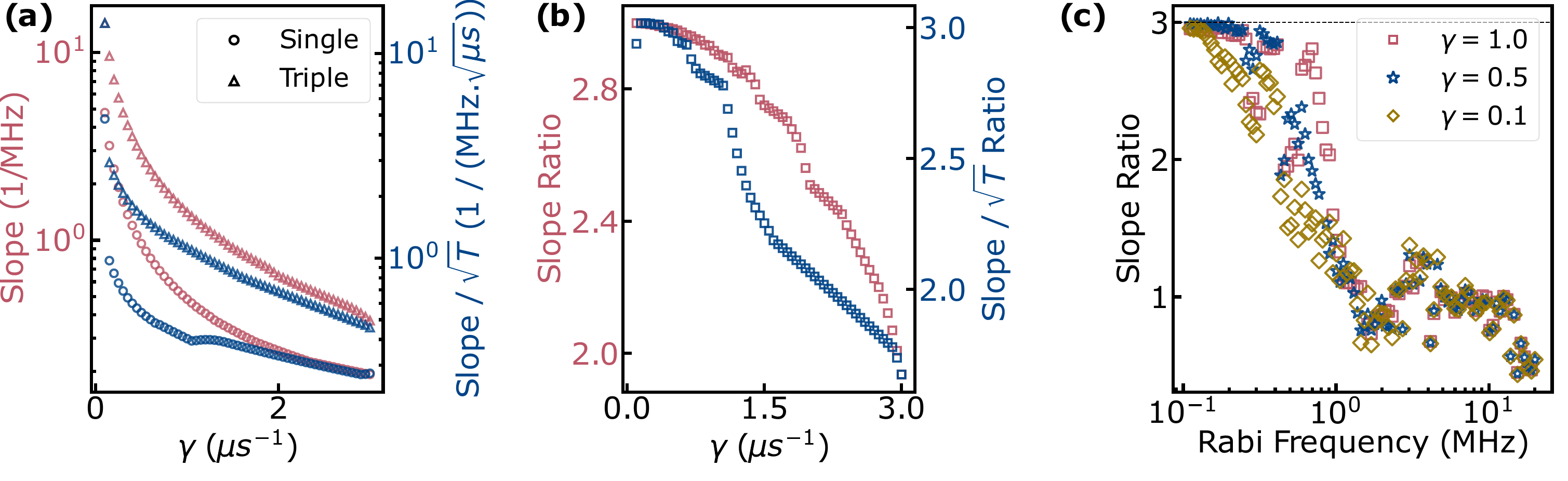}
  \caption{Simulation of pulsed ODMR sensitivity versus dephasing rate $\gamma$.  (a) Optimized slope and slope$/\sqrt{T}$ for single-tone and triple-tone. (b) Enhancement ratios (triple/single tone curves in (a)), showing a threefold improvement at low $\gamma$ with diminishing gains at higher $\gamma$. (c)Slope ratio across for fixed Rabi frequency at three dephasing rates ($\gamma = 0.1, 0.5, 1~\mu\text{s}^{-1}$)}
   \label{fig:pulsed_odmr_slope_sloperatio}
\end{figure}

To extend our analysis beyond the experimental regime, we simulated pulsed ODMR signals across a range of dephasing rates $\gamma$, optimizing our sensitivity metrics over Rabi frequency, pulse duration, and detuning at each dephasing value (Appendix~\ref{appendix:pulsed_ODMR}). Results are shown in Fig.~\ref{fig:pulsed_odmr_slope_sloperatio} (a-b). Triple-tone consistently outperforms single-tone driving, especially at low dephasing rates. At higher $\gamma$ (Appendix \ref{appendix:pulsed_ODMR}), the advantage for using triple tone diminishes. 
This is because higher Rabi frequencies are required to attain good contrast, and as the Rabi frequency approaches the hyperfine splitting, the off-resonant excitation leads to complicated interference effects. 
These results highlight that triple-tone pulsed ODMR is particularly effective in systems with long coherence times. Figure \ref{fig:pulsed_odmr_slope_sloperatio} (c) shows the dependence of slope as a function of Rabi frequency, exhibiting only a weak variation with dephasing rate, and demonstrating that triple-tone excitation is detrimental in the limit of high Rabi frequencies.



\section{Ramsey interferometry}
\label{sec:ramsey}
Ramsey interferometry is a foundational technique in NV center magnetometry~\cite{Barry_2020}. Unlike pulsed ODMR, no microwaves are applied during the free evolution interval, eliminating MW power broadening. 
Previous studies have demonstrated the advantage of triple-tone Ramsey protocols in the low-MW-power regime \cite{Arai_2022}. Here, we examine the utility of triple-tone excitation for Ramsey interferometry over a range of MW powers and dephasing rates. 

\begin{figure}[htb]
\centering
\includegraphics[width=0.8\textwidth]{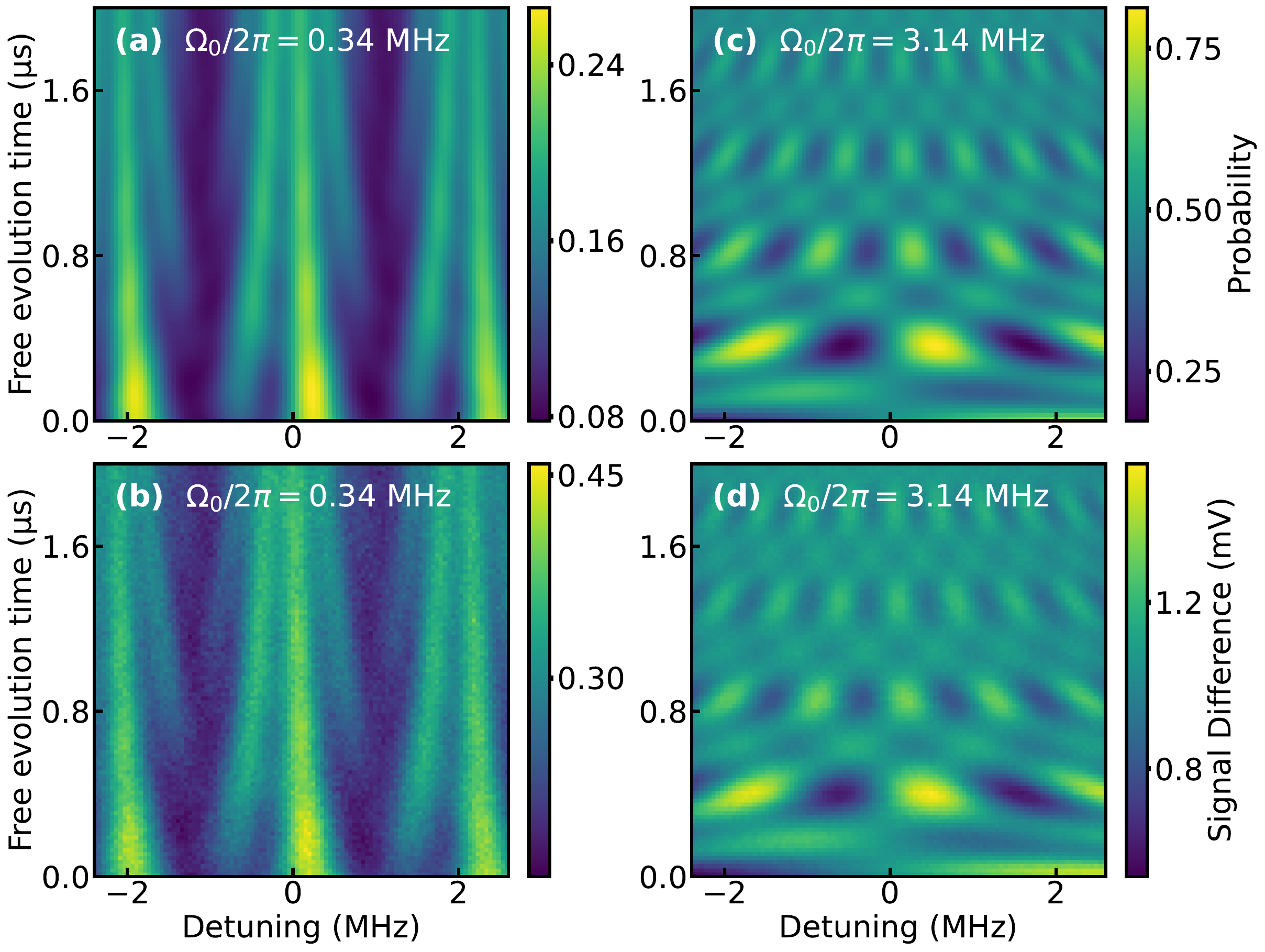}
\caption{
Single-tone Ramsey simulations (top) and corresponding experimental measurements (bottom) as a function of detuning for two Rabi frequencies. The color scales are proportional to the population in $m_s = 0$ after the pulse sequence.
(a)-(b) At 0.34 MHz, well-defined Ramsey fringes are observed, with close agreement between theory and experiment. 
(c)-(d) At 3.14 MHz, hyperfine beating distorts the fringes, accompanied by revival features from constructive interference of the $^{14}$N hyperfine components.
}

\label{fig:ramsey_combined_single}
\end{figure}

 We first establish the baseline performance for single-tone Ramsey, considering a pulse sequence of two $\pi/2$ pulses separated by a free evolution time $\tau$, with the second pulse phase-shifted by $\pi/2$. Figure~\ref{fig:ramsey_combined_single} shows experimental and simulated signals at low and high Rabi frequencies. 
 At low Rabi frequency (0.34 MHz), clear Ramsey fringes are observed, in good agreement with the simulation. At higher Rabi frequency (3.14 MHz), the signal becomes more intricate due to contributions from the three $^{14}$N hyperfine transitions, producing revival features centered at $\tau_n = n/A_{\parallel} \approx n \times 463$ ns. Slope-based analysis (Appendix~\ref{appendix:ramsey}) confirms that sensitivity is maximized at these revival points, where the hyperfine components interfere constructively.
Notably, at low Rabi frequency, where a single tone only addresses one hyperfine line, the contrast is diminished by approximately a factor of three; this corresponds to the regime where triple-tone driving is expected to enhance contrast \cite{Arai_2022}. At higher Rabi frequencies, however, full contrast is already recovered at the revival times, reducing the possible advantage of triple-tone excitation. In what follows, we quantify the transition between these regimes and evaluate the practical benefit of triple-tone excitation in NV Ramsey spectroscopy.

\begin{figure}[htb]
\centering
\includegraphics[width=1\textwidth]{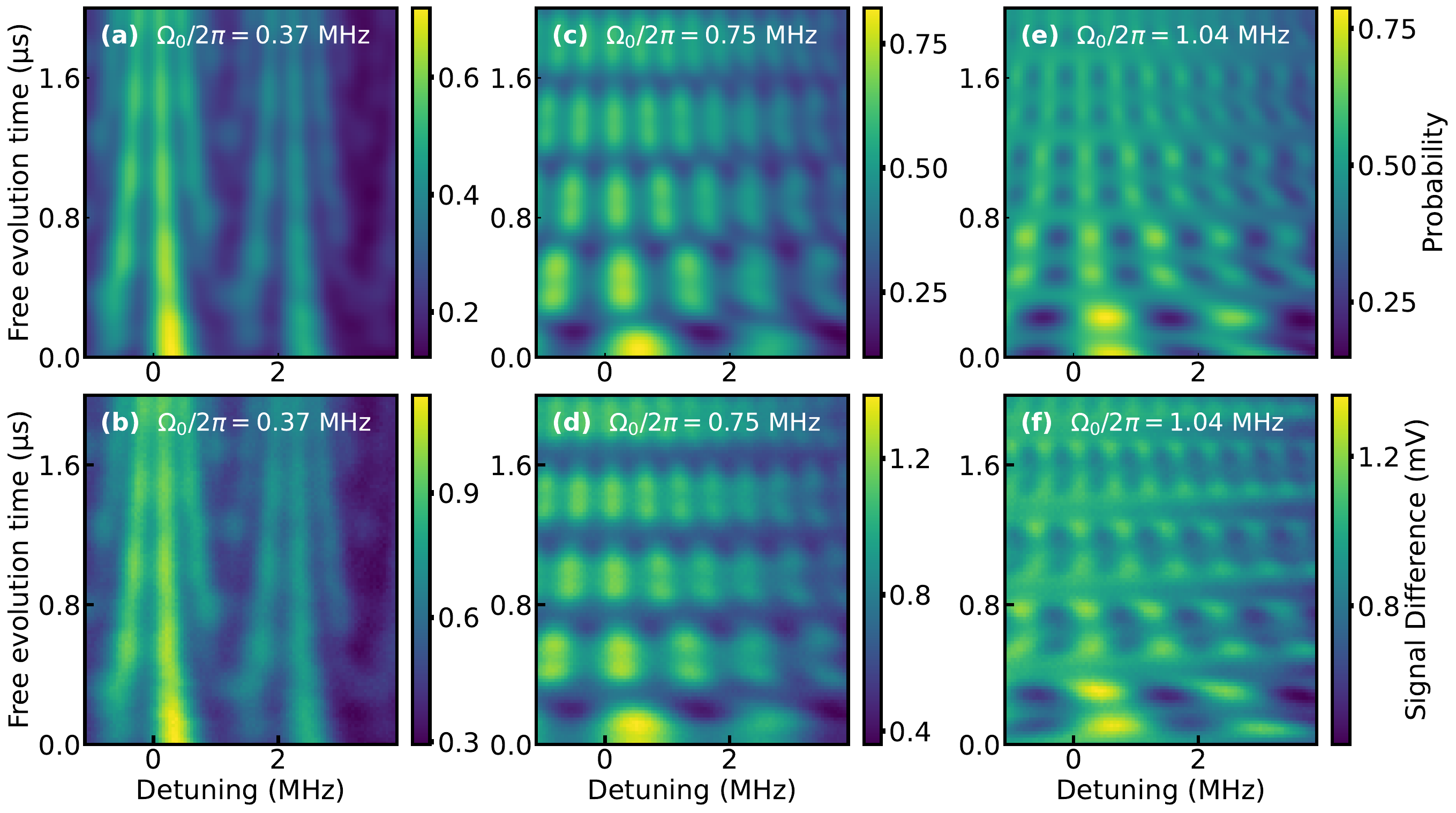}
\caption{
Triple-tone Ramsey simulations (top) and experiments (bottom) over a detuning range of –1.1 to +3.9 MHz. In each case, the signal is proportional to the population in $m_s = 0$.
(a,b) Low Rabi frequency (0.37 MHz)
(c,d) Intermediate Rabi frequency (0.75 MHz)
(e,f) Higher Rabi frequency (1.04 MHz).
}

\label{fig:ramsey_combined_triple}
\end{figure}

The triple-tone Ramsey protocol applies pulses comprised of three tones separated by the $^{14}$N hyperfine splitting, each having the same duration equal to that of a $\pi/2$ pulse for a single tone of the same Rabi frequency; each pulse sequence starts with zero relative phase between the tones, and adds a $\pi/2$ phase shift for each tone in the second pulse. Figure~\ref{fig:ramsey_combined_triple} presents both simulated and experimental data across a range of Rabi frequencies and detunings. 
At moderate Rabi frequency (0.75 MHz), the impact of off-resonant excitation becomes visible, while for Rabi frequencies > 1 MHz, we observe some discrepancies between theory and experiment that may arise from increased sensitivity to non-ideal pulse shape and hardware nonlinearities (see Appendix~\ref{appendix:ramsey}).  
The agreement at lower Rabi frequencies up to 1 MHz nevertheless validates the model for exploring regimes where triple-tone provides an advantage.

To compare the protocols, we simulate slope and slope$/\sqrt{T}$ at zero detuning over a range of dephasing rates $\gamma$, optimizing over Rabi frequency and free evolution time. We constrain the $\pi/2$ pulses to have duration $t_{\pi/2} = \pi/\sqrt{4\Omega_0^2 - \gamma^2}$, where $\Omega_0$ is the Rabi frequency and $\gamma$ is the dephasing rate. 
We also constrain the MW frequency to be on resonance and enforce a $\pi/2$ phase shift for the second pulse to avoid convergence to ODMR-like configurations (see Appendix~\ref{appendix:ramsey}). Results are shown in Fig.~\ref{fig:ramsey_slope_sloperatio}(a-b) with additional analysis in Appendix~\ref{appendix:ramsey}. 

When high Rabi frequency is possible (as allowed in our optimization), triple-tone excitation offers only a small improvement in slope at low dephasing rates and performs worse for  slope$/\sqrt{T}$, because single-tone control already achieves full contrast at the revival points. Nevertheless, when the Rabi frequency is held fixed (see Fig.~\ref{fig:ramsey_slope_sloperatio}(c)), a three-fold gain can be observed at low Rabi frequency, largely independent of the dephasing rate. We also observe oscillations in Fig.~\ref{fig:ramsey_slope_sloperatio}(c), which occur when the applied MW tone also off-resonantly drives the hyperfine transitions at effective Rabi frequencies that are even multiples of the bare Rabi frequency 
(see dashed lines and Appendix \ref{appendix:pulsed_ODMR} for more discussion). Overall, while previous triple-tone Ramsey experiments operated in the low-MW-power regime \cite{Arai_2022}, our results show that comparable sensitivity can be achieved by single-tone control at sufficiently high drive power, albeit with a more complicated signal arising from driving all three hyperfine transitions.

\begin{figure}[htb]
\centering
\includegraphics[width=1\textwidth]{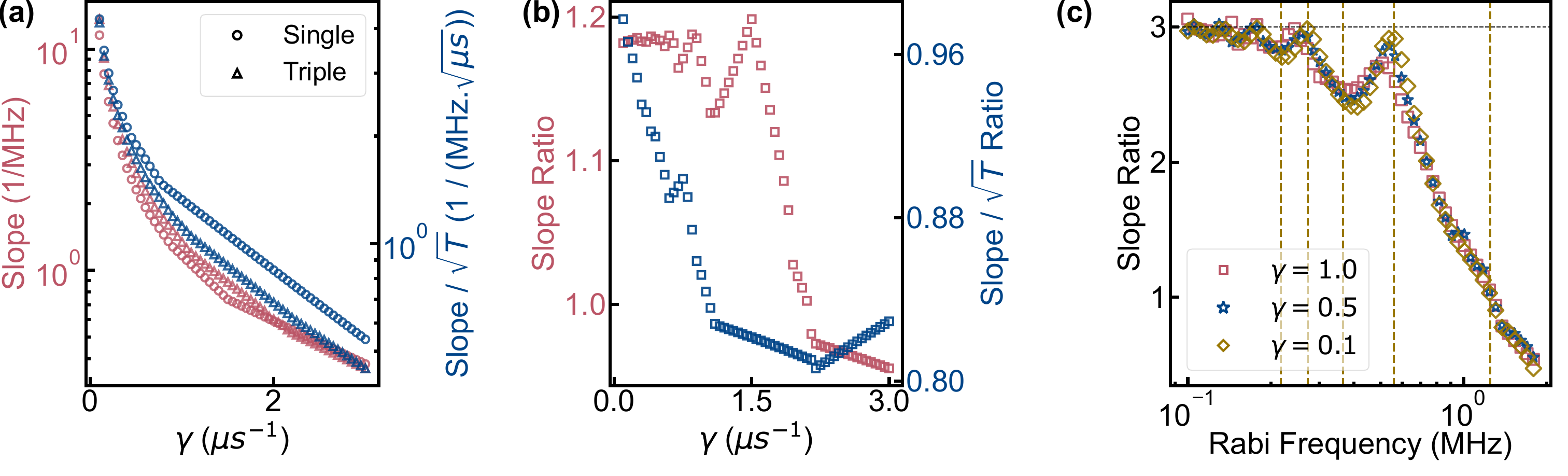}
\caption{
Simulation of Ramsey sensitivity versus dephasing rate $\gamma$. 
(a) Maximum slope and slope$/\sqrt{T}$ values (optimized over Rabi frequency and free evolution time) plotted on a logarithmic scale for single-tone and triple-tone excitation.
(b) Corresponding ratios of triple-to-single-tone sensitivity, showing modest triple-tone improvements, primarily at low dephasing rates.
(c) Slope ratio across for fixed Rabi frequency at three dephasing rates ($\gamma = 0.1, 0.5, 1~\mu\text{s}^{-1}$). Dashed lines show Rabi frequencies satisfying $\sqrt{\Omega_0^2 + A_\parallel^2} = 2n\Omega_0$ with $n$ an integer.
}
\label{fig:ramsey_slope_sloperatio}
\end{figure}

\section{Discussion}
\label{sec:discussion}
In this work, we compared single-tone and triple-tone microwave pulse sequences for pulsed ODMR and Ramsey interferometry in NV ensemble magnetometry. Using simulations validated by experiment, we quantified how these control schemes affect sensitivity under varying Rabi frequencies and dephasing rates, with the aim of quantifying the regimes where triple-tone control could offer meaningful advantages. 

For pulsed ODMR, triple-tone driving enables simultaneous excitation of all three $^{14}$N hyperfine transitions, yielding up to a threefold enhancement in signal slope for low-dephasing samples. This provides a practical route to improving sensitivity without increasing sequence complexity.

Ramsey interferometry, however, presents a more nuanced picture. While triple-tone driving 
can coherently address all hyperfine components, 
its performance is highly regime-dependent. Here, Rabi frequency plays a more dominant role than dephasing rates. At low Rabi frequencies, we observed sensitivity improvements similar to those seen in pulsed ODMR, arising from enhanced contrast. As the Rabi frequency increases, the three MW tones begin to effectively drive all three hyperfine transitions, leading to interference effects that degrade sensitivity. On the other hand, single-tone Ramsey becomes ever more sensitive with increasing Rabi frequency, such that triple-tone excitation affords little to no sensitivity advantage when high Rabi frequencies are attainable. 
However, there may be other considerations: While low-power triple-tone and high-power single-tone Ramsey protocols offer similar sensitivity, low-power triple-tone produces sinusoidal signals without hyperfine beating that may be advantageous for straightforward data interpretation as well as application of adaptive \cite{McMichael2021, PhysRevApplied.18.054001, Kelley2025, Zohar_2023} or non-adaptive \cite{PhysRevB.90.024422, Waldherr2012} phase estimation techniques.  


\section{Conclusion}
\label{sec:conclusion}
Our results explore the regimes where multi-frequency driving can enhance NV ensemble magnetometry, showing that its effectiveness strongly depends on the protocol. For pulsed ODMR, triple-tone driving achieves up to a threefold enhancement in sensitivity, making it a practical and effective route to boost performance in compact NV magnetometers. In contrast,  triple-tone control boosts the sensitivity of Ramsey interferometry only in the low-MW-power regime. At higher Rabi frequencies, off-resonant excitation limits practical gains for triple-tone leaving single-tone Ramsey as the more reliable choice for applications where high Rabi frequencies are accessible.

Beyond triple-tone control, several strategies remain open for further mitigating the impact of hyperfine structure on NV ensemble magnetometry. Nuclear spin polarization, for example, can improve initialization fidelity by polarizing the $^{14}$N or nearby $^{13}$C nuclear spins, reducing decoherence and enhancing contrast, though this typically requires additional hardware \cite{Busaite_2020}, and functions poorly with high dephasing rates. Extending multi-tone protocols to NV centers with ${}^{15}$N \cite{B_rgler_2023,PhysRevB.80.241204,Oon_2022}—which feature a simpler two-level hyperfine structure—may preserve constructive benefits of multi-frequency control while reducing interference effects. By mapping out the sensitivity gain associated with multi-frequency driving, our results can aid researchers in choosing the best strategy to combat the contrast loss associated with hyperfine interactions in their experiment-specific regime. 


\section*{Back matter}
\begin{backmatter}
\bmsection{Funding}
 This work was carried out as a part of the Quantum Sensors Challenges Program (QSP) and supported by the National Research Council (NRC) grant QSP-051. 


\bmsection{Disclosures}
Disclosure code E: LC was formerly employed by SBQuantum from Feb. 15, 2024- Feb. 14, 2025. Disclosure code F: LC discloses research funding through NRC QSP-051, which includes in-kind contributions from SBQuantum.
YBL declares no conflicts of interest.
DRG and VH hold equity in SBQuantum.

\bmsection{Data availability} Data underlying the results presented in this paper are not publicly available at this time but may be obtained from the authors upon reasonable request.

\end{backmatter}


\appendix

\section*{Appendix}

\section{Master equations for simulation}
\label{appendix:theory}
The master equation for a spin-$1/2$ system driven by single-frequency MW can be represented in the rotating frame and rotating-wave approximation by:

\begin{equation}
\frac{d}{dt} 
\begin{pmatrix} 
\rho_{11}(t) \\ 
\rho_{12}(t) \\ 
\rho_{21}(t) \\ 
\rho_{22}(t) 
\end{pmatrix} 
=
\begin{pmatrix}
0 & -\frac{i}{2}\Omega & \frac{i}{2}\Omega^* & \Gamma \\
-\frac{i}{2}\Omega^* & -\gamma - i\delta & 0 & \frac{i}{2}\Omega^* \\
\frac{i}{2}\Omega & 0 & -\gamma + i\delta & -\frac{i}{2}\Omega \\
0 & \frac{i}{2}\Omega & -\frac{i}{2}\Omega^* & -\Gamma
\end{pmatrix}
\begin{pmatrix} 
\rho_{11}(t) \\ 
\rho_{12}(t) \\ 
\rho_{21}(t) \\ 
\rho_{22}(t) 
\end{pmatrix}
\label{eq:bloch_eqn_general}
\end{equation}
where $\rho_{11} (t)$ and $\rho_{22} (t)$ are the populations of the two sublevels of the system, $\rho_{12} (t)$ and $\rho_{21} (t)$ are the coherences, $\Omega$ is the driving Rabi frequency, $\Gamma = 1/{T_1}$ is the spin-lattice relaxation rate, $\gamma = 1/{T_2^*}$ is the effective ensemble dephasing rate and $\delta$ is the MW detuning with respect to the resonance frequency of the targeted NV orientation.

With a triple-tone drive, $\Omega$ is time-dependent:
\begin{equation}
\Omega(t) = \Omega_0 e^{i \phi} \left(1 + \epsilon e^{i( \omega t + \xi_1)} + \epsilon e^{-i( \omega 
 t + \xi_2 )}\right)
\label{eq:new_Omega},
\end{equation}
where $\omega = 2\pi A_{\parallel}$ is the detuning of each sideband tone from the carrier, $\phi$ is the global MW phase, $\xi_i$ is the relative phase of tone $i$, and $\epsilon$ controls the sideband strength. 
$\epsilon = 0$ corresponds to a single MW drive and  $\epsilon = 1$ results in a triple-tone MW drive. Substituting Eq.~\ref{eq:new_Omega} in Eq.~\ref{eq:bloch_eqn_general}, the new set of equations to model triple-tone-driven dynamics is given by:

\begin{equation}
\frac{d}{dt}
\begin{pmatrix}
\rho_{11}(t) \\
\rho_{12}(t) \\
\rho_{21}(t) \\
\rho_{22}(t)
\end{pmatrix}
=
\begin{pmatrix}
0 & -\frac{i}{2} C_{12}(t)\Omega_0 & \frac{i}{2} C_{21}(t)\Omega_0 & \Gamma \\
-\frac{i}{2} C_{21}(t)\Omega_0 & -\gamma - i \delta & 0 & \frac{i}{2} C_{21}(t) \Omega_0\\
\frac{i}{2} C_{12}(t)\Omega_0 & 0 & -\gamma + i \delta & -\frac{i}{2} C_{12}(t)\Omega_0 \\
0 & \frac{i}{2} C_{12}(t)\Omega_0 & -\frac{i}{2} C_{21}(t)\Omega_0 & -\Gamma
\end{pmatrix}
\begin{pmatrix}
\rho_{11}(t) \\
\rho_{12}(t) \\
\rho_{21}(t) \\
\rho_{22}(t)
\end{pmatrix}
\label{eq:bloch_eqn_tt}
\end{equation}
where 
$C_{12}(t) = e^{i \phi} \left(1 + \epsilon e^{i( \omega t + \xi_1)} + \epsilon e^{-i( \omega 
 t + \xi_2 )}\right)$ and $C_{21} = C_{12}^*$. 

For simulations, the system is initialized in $\rho_{11}(0)=1$, with $\xi_i=0$, and $\epsilon=0$ or 1 for single or triple-tone driving respectively. For Pulsed ODMR, we optimize for the $\pi$ pulse duration to maximize the contrast, while Ramsey is modeled such that it uses two $\pi/2$ pulses of duration $t_{\pi/2} = \pi/\sqrt{4\Omega_0^2 - \gamma^2}$ separated by a free evolution time $\tau$, with the second pulse phase-shifted by $\phi=\pi/2$. To account for the phase accumulated during free evolution and the first $\pi/2$ pulse, for the second pulse we set $\xi_i \mapsto \xi_i + \omega(\tau+t_{\pi/2})$. Hyperfine structure is included in the secular approximation by averaging the signal over detunings $\delta=\delta_0-2\pi A_{\parallel},\;\delta_0,\;\delta_0+2\pi A_{\parallel}$. Sensitivity metrics are obtained from the derivative of $\rho_{22}(t)$ with respect to detuning, evaluated at the end of the sequence.

Our model approximates the NV spin system as an effective two-level system, rather than employing the full $9\times 9$ NV–$^{14}$N Hamiltonian. We make this approximation because, at the moderate magnetic fields we consider, 
the transitions from $m_s = 0$ to $m_s = -1$ and from $m_s = 0$ to $m_s = +1$ are split by a frequency far exceeding the Rabi frequency, such that off-resonant excitation of the other transition is negligible.  In these moderate magnetic fields, the off-axis hyperfine couplings remain strongly suppressed by the zero field splitting, yielding corrections of order $A_{\perp}^2/\Delta^2$ where $A_\perp \approx 2.7$ MHz \cite{PhysRevB.79.075203}. We also consider the applied magnetic field to be well aligned with the NV axis and sufficiently far from level anticrossings, so that nuclear-spin mixing and dynamic nuclear polarization are negligible. The model assumes ideal square microwave pulses and does not capture distortions in pulse amplitude or phase that can arise when driving multiple tones simultaneously due to dispersion in the microwave circuitry. Similarly, variations across the NV ensemble, such as magnetic-field gradients or local strain differences, are assumed to be minor compared to the hyperfine splitting. Within these constraints, the effective two-level description enables efficient parameter sweeps over Rabi frequency, pulse length, and dephasing while maintaining good quantitative agreement with experiment. A full $9\times 9$ treatment would be required to capture dynamical nuclear effects due to strong off-axis fields, and electron–nuclear entanglement, which are beyond the scope of this work.

\section{Pulsed ODMR Sensitivity Optimization}
\label{appendix:pulsed_ODMR}
\begin{figure}[htb]
  \centering
    \includegraphics[width=0.6\textwidth]{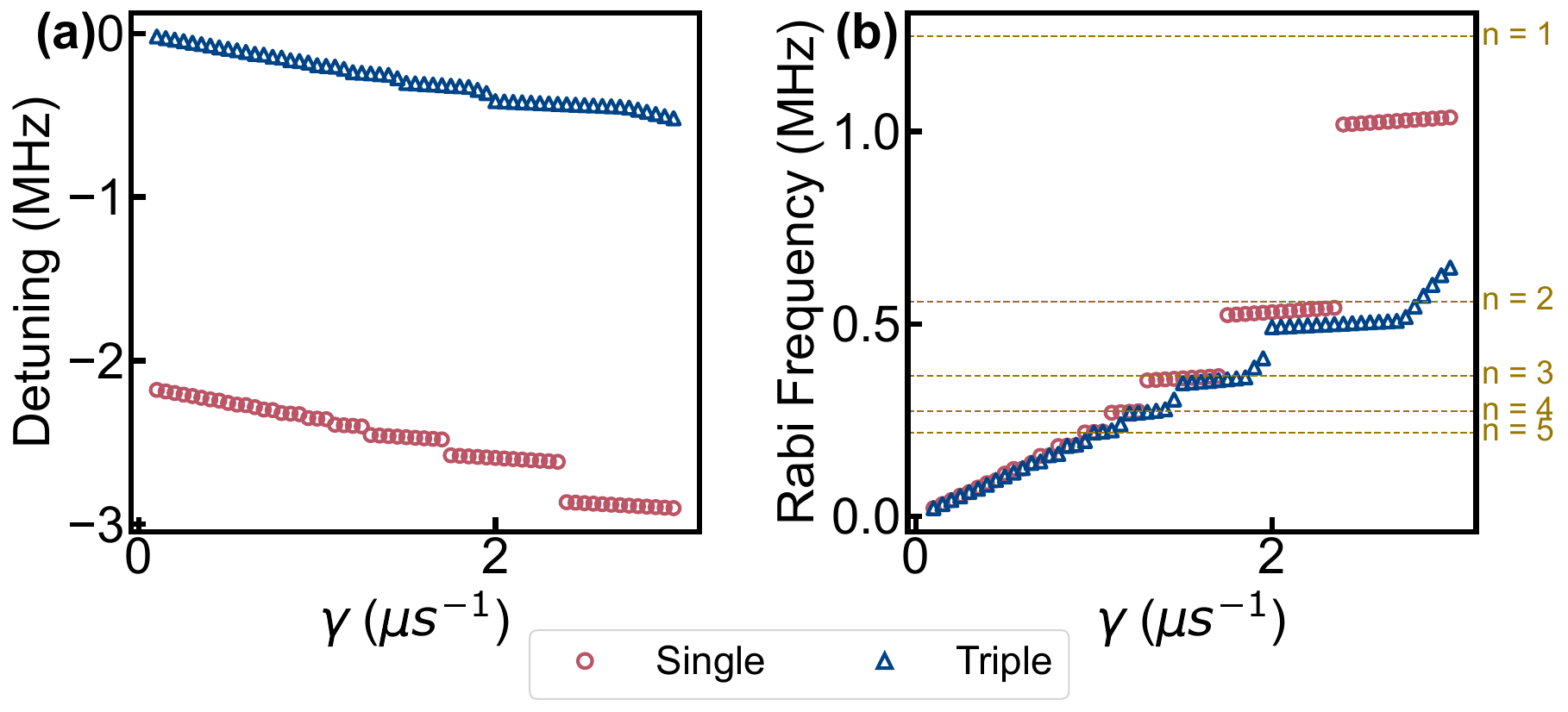}
    \caption{
    Numerically optimized parameters for maximizing pulsed ODMR sensitivity (slope metric) as a function of dephasing rate $\gamma$. 
    (a) Optimal detuning: single-tone control favors side fringes, while triple-tone driving shifts the optimum to near-zero detuning. 
    (b) Corresponding optimal Rabi frequencies, which cluster around integer multiples of the $\pi$-pulse frequency (dashed lines).
    }
    \label{fig:pulsedODMR_appendix}
\end{figure}

To determine the maximum slope and slope$/\sqrt{T}$ sensitivity metrics for each dephasing rate $\gamma$, we used Mathematica’s \texttt{DifferentialEvolution} algorithm to optimize with respect to pulse duration, detuning, and Rabi frequency. Figure~\ref{fig:pulsedODMR_appendix}(a) shows the detuning values that maximize the slope. For single-tone control, the optimum lies at the side fringes, where overlapping resonance tails produce the steepest slope. In contrast, triple-tone driving benefits from the same effect but achieves an even steeper slope at the central fringe due to the combined contrast enhancement. The same trend is observed for slope$/\sqrt{T}$. The overall trend towards greater detuning with dephasing rate can be explained by line-broadening, which shifts the location of maximal slope to larger detuning.

Figure~\ref{fig:pulsedODMR_appendix}(b) shows the corresponding optimal Rabi frequencies. 
Perhaps surprisingly, the optimal Rabi frequency does not change smoothly with dephasing rate. This behavior can be understood qualitatively by recognizing that when a MW tone is near-resonant with one of the hyperfine transitions, with a 
Rabi frequency $\sim \Omega_0$, it also off-resonantly drives other hyperfine transitions that are detuned by $\sim A_{\parallel}$ with an effective Rabi frequency 
$\sim \sqrt{\Omega_0^2 + (2\pi A_{\parallel})^2}$. The MW tone will be most selective in driving its near-resonant transition when a $\pi$ pulse on the resonant transition corresponds to a $2n\pi$ pulse on the off-resonant transitions. The Rabi frequencies that accomplish this task are illustrated by dotted lines in 
Fig.~\ref{fig:pulsedODMR_appendix}(b). The optimal Rabi frequencies do not lie precisely on these lines because the optimal detuning is not precisely on resonance with one of the hyperfine lines (particularly at higher dephasing rates). Nevertheless, the good agreement at low dephasing rates and the overall trend in the steps between optimal Rabi frequencies indicate that this qualitative mechanism underlies our observations.

\section{Ramsey}
\label{appendix:ramsey}
\subsection{Ramsey Simulation Analysis}

\begin{figure}[htb]
  \centering
    \includegraphics[width=1\textwidth]{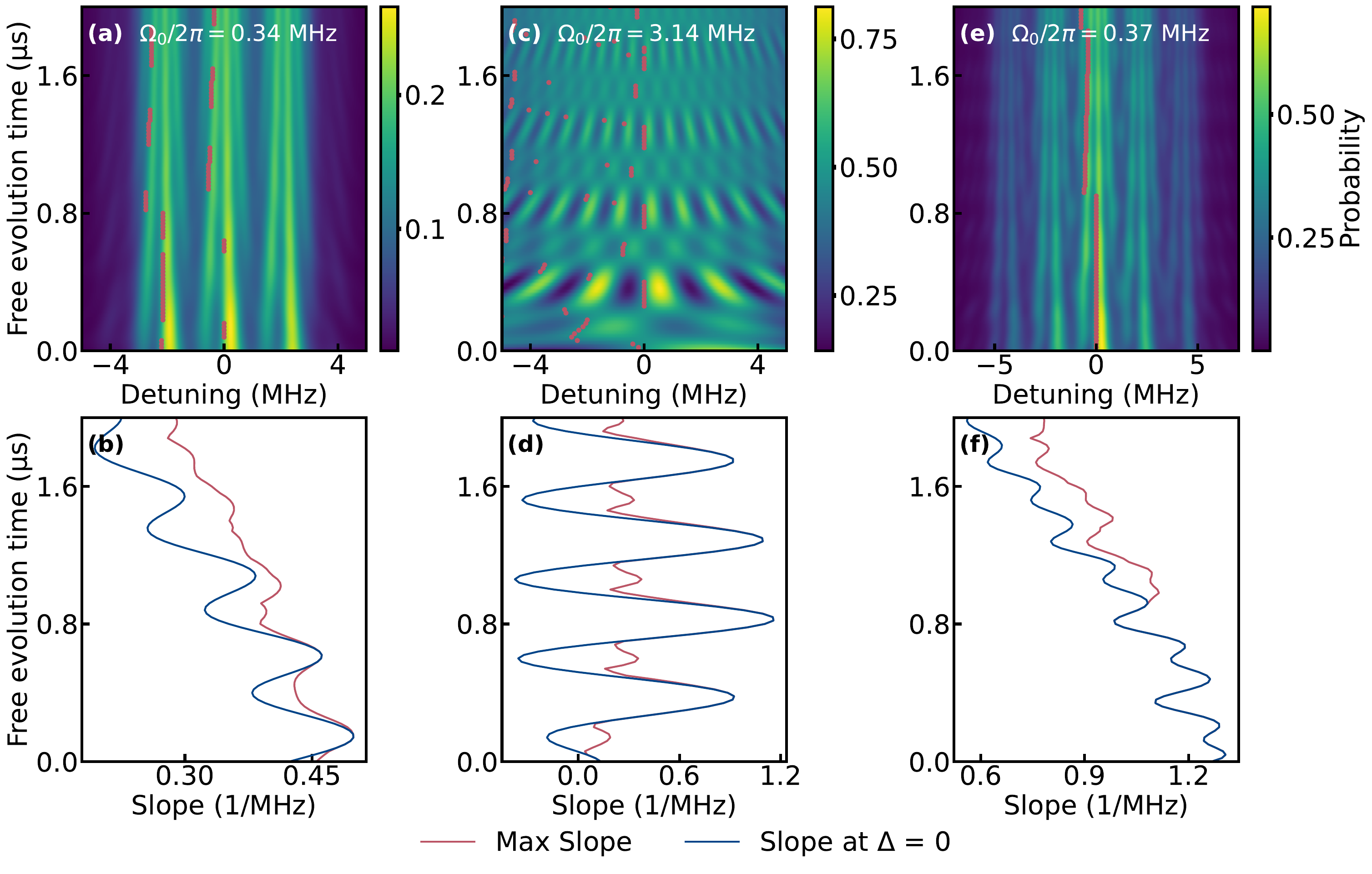}
   \caption{Ramsey simulations. (a,c,e) Spin-flip probability as a function of detuning and free evolution time for single-tone at $\Omega_0/(2\pi)=0.34$~MHz (a), single-tone at $\Omega_0/(2\pi)=3.14$~MHz (c), and triple-tone at $\Omega_0/(2\pi)=0.37$~MHz (e). (b,d,f) Comparison between the slope value at zero detuning (blue curves) and the global maximum (red curves obtained from the red markers appearing in (a,c,e)). 
   }
   \label{fig:ramsey_appendix_A}
\end{figure}

All Ramsey simulations were performed with step sizes of 20~ns for the free evolution time $\tau$ and 0.01~MHz for the detuning $\delta$. We directly calculated both the population remaining in $m_s = 0$ and its derivative with respect to detuning by solving coupled differential equations for the two quantities, such that finite step sizes do not affect estimates of the slope. In the main text, we constrained the detuning to zero throughout because 
we have introduced a relative $\pi/2$ phase shift between the two microwave pulses, such that zero detuning is optimal for a two-level system; all sensitivities reported in the main text are evaluated at $\delta = 0$. Figure~\ref{fig:ramsey_appendix_A} 
justifies this choice: Although the maximum slope can occur at a non-zero detuning, 
for the free evolution times that yield the best
sensitivity, the maximum slope is negligibly different from the maximum slope at zero detuning. At low Rabi frequency, the difference between the red and blue curves in Fig.~\ref{fig:ramsey_appendix_A}(b) is most likely related to the sequence acquiring ODMR-like character at low Rabi frequency, where the overall change in contrast with detuning (due to imperfect driving off resonance) pushes the maximum slope away from $\delta = 0$. 
At high Rabi frequency, for free evolution times away from the hyperfine revivals, interference between the three hyperfine transitions can favor a non-zero detuning. 
Nevertheless, near the optimal-sensitivity revival times, the maximum slope is consistently found at zero detuning. For this reason, and to reduce the dimensionality of the optimization, we fix the detuning to zero in the subsequent analysis. 

\begin{figure}[H]
  \centering
    \includegraphics[width=1\textwidth]{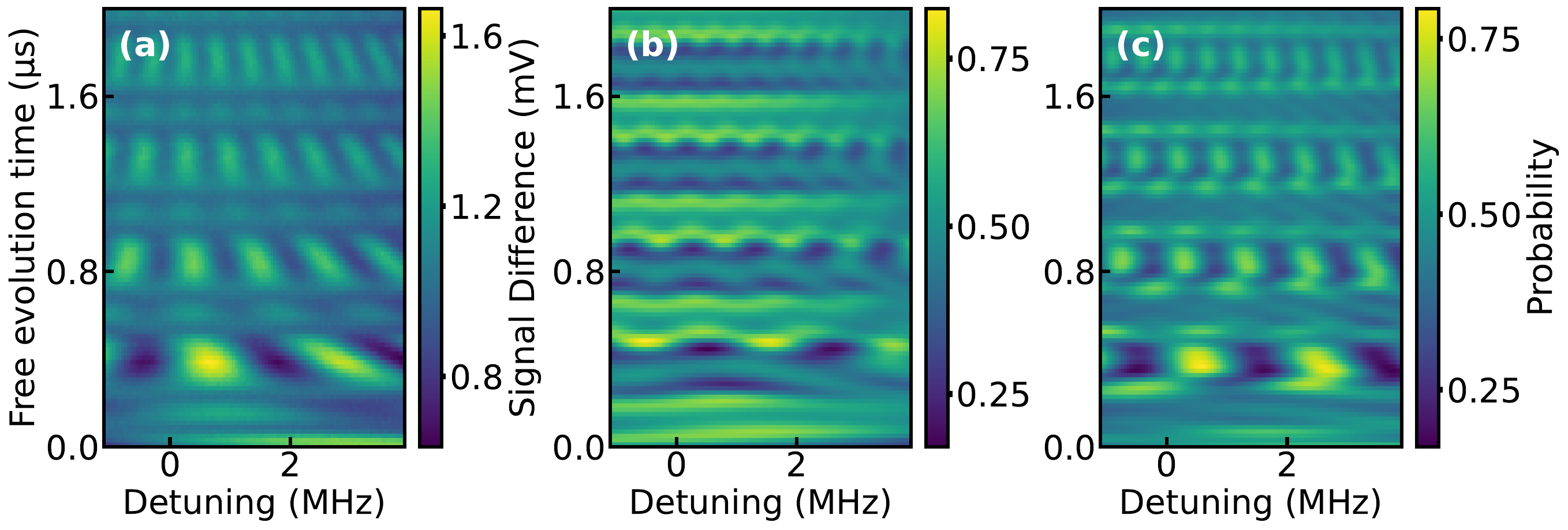}
  \caption{Triple-tone Ramsey simulations at $\Omega_0/(2\pi)=3.16$~MHz. 
  (a) Experimental data. (b) Idealized simulation with no over-rotation and $\xi_i = 0$. (c) Simulation including $\pi/2$ pulse over-rotation of 1.4$\times$ and relative phase offsets ($\xi_1 = 17\pi/25$, $\xi_2 = 18\pi/25$) between hyperfine tones. 
}
 \label{fig:ramsey_appendix_triple_high}
\end{figure}

While our simulations match well with the experiment at low-to-moderate Rabi frequencies, we observe some discrepancies for high Rabi frequency ($\Omega_0/(2\pi)\approx$3.16~MHz) triple-tone experiments (see Fig.~\ref{fig:ramsey_appendix_triple_high}). 
Possible reasons for the discrepancies include: (i) over-rotation in the experimental $\pi/2$ pulses, absent in the idealized simulation, and (ii) non-zero initial phases $\xi_i$ between hyperfine tones due to IQ mixer imperfections. Incorporating both over-rotation and relative phase offsets in the simulation yields much closer agreement with experiment (see Fig.~\ref{fig:ramsey_appendix_triple_high}(c)), suggesting the need for more precise pulse calibration. Additionally, our model assumes uniform Rabi frequency across the ensemble, neglecting spatial inhomogeneities that may become more pronounced at high drive amplitudes and contribute to the complex interference observed experimentally.

\subsection{Sensitivity Optimization}
\begin{figure}[htb]
  \centering
    \includegraphics[width=0.8\textwidth]{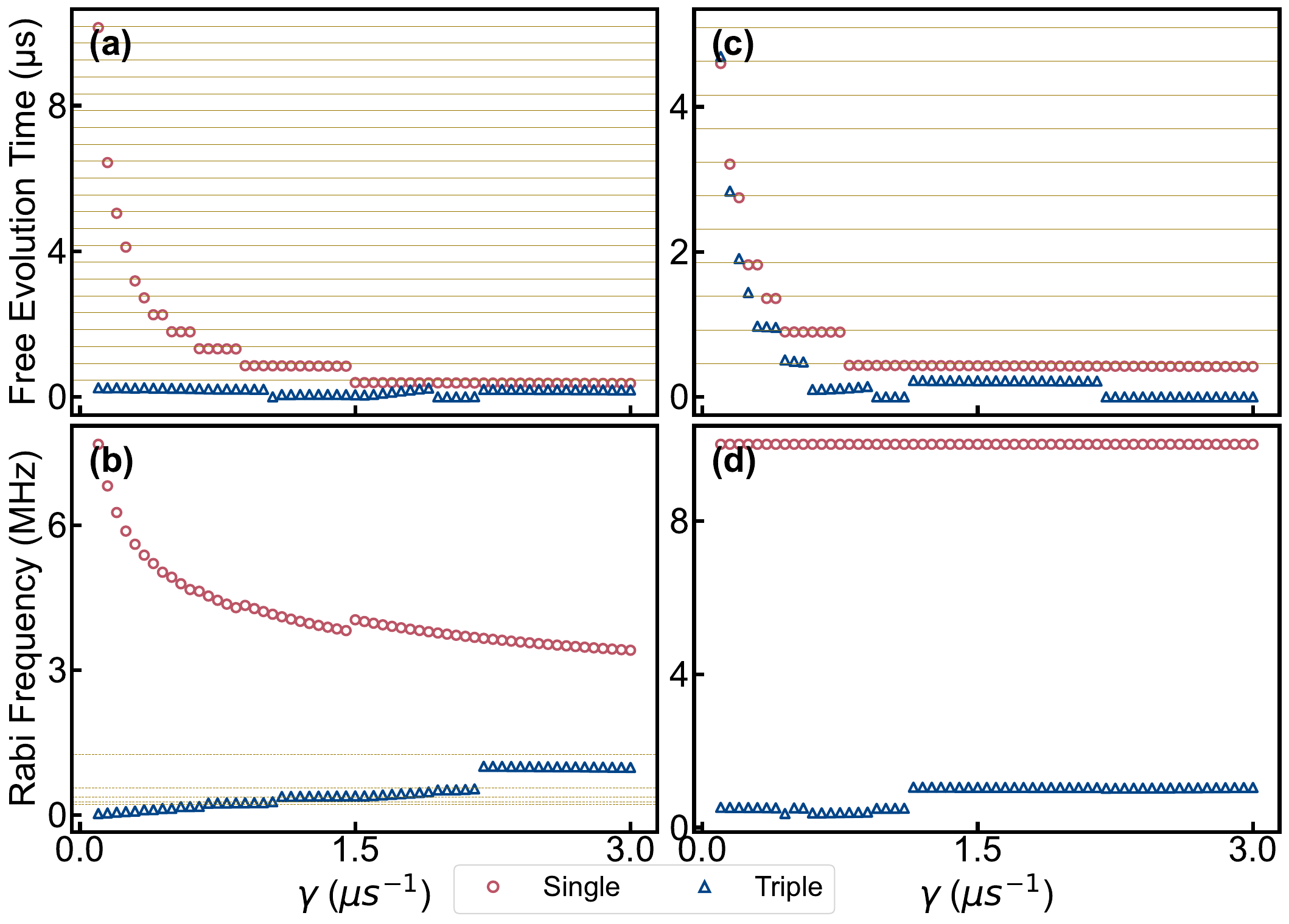}
    \caption{Parameter settings for optimal single- and triple-tone Ramsey sensitivity metrics (as shown in Fig.~\ref{fig:ramsey_slope_sloperatio}(a-b)).  (a,b) Free evolution time and Rabi frequency that maximize slope. (c,d) Corresponding results for slope$/\sqrt{T}$. Horizontal lines in (a,c) represent hyperfine revival times. In (b), horizontal lines correspond to the same Rabi frequencies shown in Fig.~\ref{fig:pulsedODMR_appendix}(b).
  }
   \label{fig:ramsey_appendix_B}
\end{figure}

The \texttt{DifferentialEvolution} method in Mathematica was used for optimization of both slope and slope$/\sqrt{T}$ metrics. The optimization parameters were the free evolution time $\tau$ and the Rabi frequency $\Omega_0$, while the detuning $\delta$ was fixed to zero. The duration of each $\pi/2$ pulse was set to 
$t_{\pi/2} = \pi/\sqrt{4\Omega_0^2-\gamma^2},$
corresponding to half of the pulse duration that achieves maximal spin flip probability in the presence of dephasing for a two-level system. 
Figure~\ref{fig:ramsey_appendix_B}(a,c) shows the trend in optimal free evolution time with $\gamma$. The steps in optimal free evolution time are approximately consistent with hyperfine revival times; the agreement is not exact because the $\pi/2$ pulses 
have finite duration. Interestingly, for triple-tone configurations, the optimization sometimes favors $t = 0$, effectively reducing the protocol to pulsed ODMR with a mid-pulse phase shift. Figure~\ref{fig:ramsey_appendix_B} shows that optimal Rabi frequencies for maximizing slope tend to form discrete steps, corresponding qualitatively to steps observed in pulsed ODMR, while free evolution durations favor the hyperfine revival times. 

\begin{figure}[H]
  \centering
    \includegraphics[width=1\textwidth]{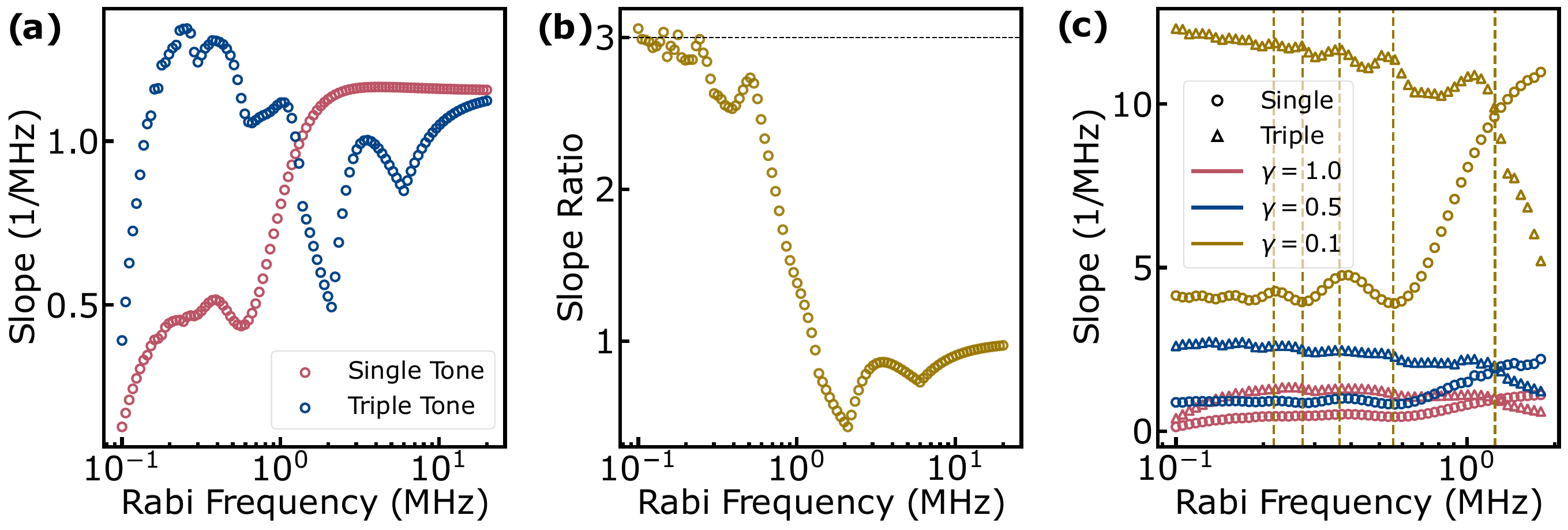}
    \caption{Dependence of Ramsey sensitivity (slope metric) on Rabi frequency.  
    (a–b) Comparison between single-tone and triple-tone control at a fixed dephasing rate $\gamma = 1~\mu\text{s}^{-1}$, showing enhanced performance for triple-tone at low Rabi frequencies. At higher Rabi frequencies, both control schemes converge.  Note that (b) is equivalent to Fig.~\ref{fig:ramsey_slope_sloperatio}(c) over a wider range of Rabi frequencies.
    (c) Slope dependence on Rabi frequency over a restricted frequency range and across multiple dephasing rates ($\gamma = 0.1, 0.5, 1~\mu\text{s}^{-1}$), illustrating the slope values from which the ratios shown in Fig.~\ref{fig:ramsey_slope_sloperatio}(c) are taken. 
    }
 \label{fig:ramsey_slope_sloperatio_dephasing_1_all}
\end{figure}

For slope$/\sqrt{T}$, single-tone control driven at high power (here limited to 10~MHz in simulations in Fig. \ref{fig:ramsey_appendix_B}(d)) can match or even outperform triple-tone, as the $1/\sqrt{T}$ factor leads to additional advantage at the high Rabi frequencies where single-tone Ramsey realizes the highest slope. This trend is further illustrated in Fig.~\ref{fig:ramsey_slope_sloperatio_dephasing_1_all}, 
which shows the dependence of the single- and triple-tone slope sensitivity metrics on the Rabi frequency. At low drive strengths, triple-tone offers a clear advantage, but this benefit diminishes as $\Omega_0$ increases. At sufficiently high Rabi frequencies, single-tone and triple-tone achieve nearly equivalent performance, and, as shown in the main text Fig.~\ref{fig:ramsey_slope_sloperatio}(c), this behavior remains consistent across different dephasing rates. The oscillations in Fig. \ref{fig:ramsey_appendix_B}(c) correspond to the same values as observed in pulsed ODMR (Fig. \ref{fig:pulsedODMR_appendix}(b)).

\bibliography{sample}


\end{document}